\documentclass[11pt]{article}
\usepackage{moriond,epsfig}

\def\Journal#1#2#3#4{{#1} {\bf #2}, #3 (#4)}

\def\JETP{\em Journ. Exp. Theor. Phys.}
\def\JPG{{\em Journ. Phys.} G: {\em Nucl. Part. Phys.}}
\def\APP{\em Astropart. Phys}

\def\be{\begin{equation}}
\def\ee{\end{equation}}
\def\bea{\begin{eqnarray}}
\def\eea{\end{eqnarray}}

\begin{document}
\vspace*{4cm}
\title{THE ORIGIN OF THE KNEE IN THE COSMIC-RAY ENERGY SPECTRUM}
\author{\underline{A.D. ERLYKIN}$^{1,2}$, A.W.WOLFENDALE$^2$ }
\address{$^1$ P.N.Lebedev Physical Institute, 
 Leninsky pr. 53, Moscow 117924, Russia \\ 
 $^2$ Physics Department, University of Durham, Durham DH1 3LE, UK } 

\maketitle\abstracts
{A sudden steepening of the cosmic-ray energy spectrum ( the knee ) is
observed at an energy of about 3 PeV (1 PeV = 10$^{15}$eV). The recent
results on extensive air showers allow us to conclude that: a) the knee has
an astrophysical origin; b) the `sharpness' and the fine structure of
the knee rule out `Galactic Modulation' as the origin of the knee;
c) most likely the knee is the result of the explosion of a single,
recent, nearby supernova.}
\section{Introduction}

Cosmic rays spread over 11 decades of energy with an almost
featureless power law spectrum. There are just two structures which
are well established:
the steepening at an energy of about $3 \cdot 10^{6}$ GeV and the flattening
near $10^{10}$ GeV. The first is called {\em the knee}, the second is
{\em the ankle}. Both features are crucial for understanding 
cosmic-ray origin and propagation. We concentrate on the knee origin,
because there has been progress in its study during the last few years. 

\section{Models for the Origin of the Knee}

Models of the knee proposed to date can be divided into two distinct
classes: {\em astrophysical} and {\em interaction} models. The 
{\em astrophysical} models attribute the change in the spectra of the 
observed extensive air showers (~EAS~)
to the change in the energy spectra of the primary cosmic
rays. The {\em interaction} models imply that the primary energy spectrum
has no such sharp change and the observed steepening of EAS size
spectra is due to the sudden change of the nature of the interactions 
between the high energy particles of primary cosmic rays and the
atmosphere. The astrophysical models are more numerous and
developed. They might be also subdivided into two classes: the {\em source}
models, with a change of sources or their acceleration mechanisms, and
the {\em propagation} models, with a change of the cosmic-ray
propagation between the source and the observer. 

\section{EAS Characteristics in the Knee Region}
The basic characteristics which are important for the analysis of
the knee origin are the energy spectrum, the mass composition of
the primary particles and the anisotropy of their arrival directions.
The first two determine the spectral shape and ratios of
different EAS components. 

\subsection{Shape of the EAS size spectra}

The measurements of EAS clearly show \cite{ew2} that:

$*$ the size spectra of {\em all} components have a sharp knee at the values
corresponding to a primary energy of about 3 PeV.

This is an important result because {\em it does not leave room for the 
interaction models}.  

$*$ the EAS electron size corresponding to the knee position decreases
with the atmospheric depth in accordance
with expectation, if the knee occurs at a fixed primary energy.    

$*$ the sharpness of the knee (~$1.3\pm0.1$~) exceeds that expected in
the Galactic Modulation model (~$\sim 0.3$~). 

$*$ the knee has the fine structure. If all the electron size spectra are
normalized at their knee position, 
then, at the size which is by the factor of 4 higher, there is another 
intensity peak. The second peak is
found also in Cherenkov light spectra, which proves its astrophysical origin. 
These results are shown in Figure 1. Here, the fine structure of the 40 EAS
size and 5 Cherenkov light spectra are shown for the excess of the
intensity over the running mean 
\cite{ew2}. The presence of the second peak
is seen beyond the error bars both in EAS size and in
Cherenkov light spectra.
\begin{figure}[htb]
\begin{center}
\includegraphics[height=7cm,width=15cm]{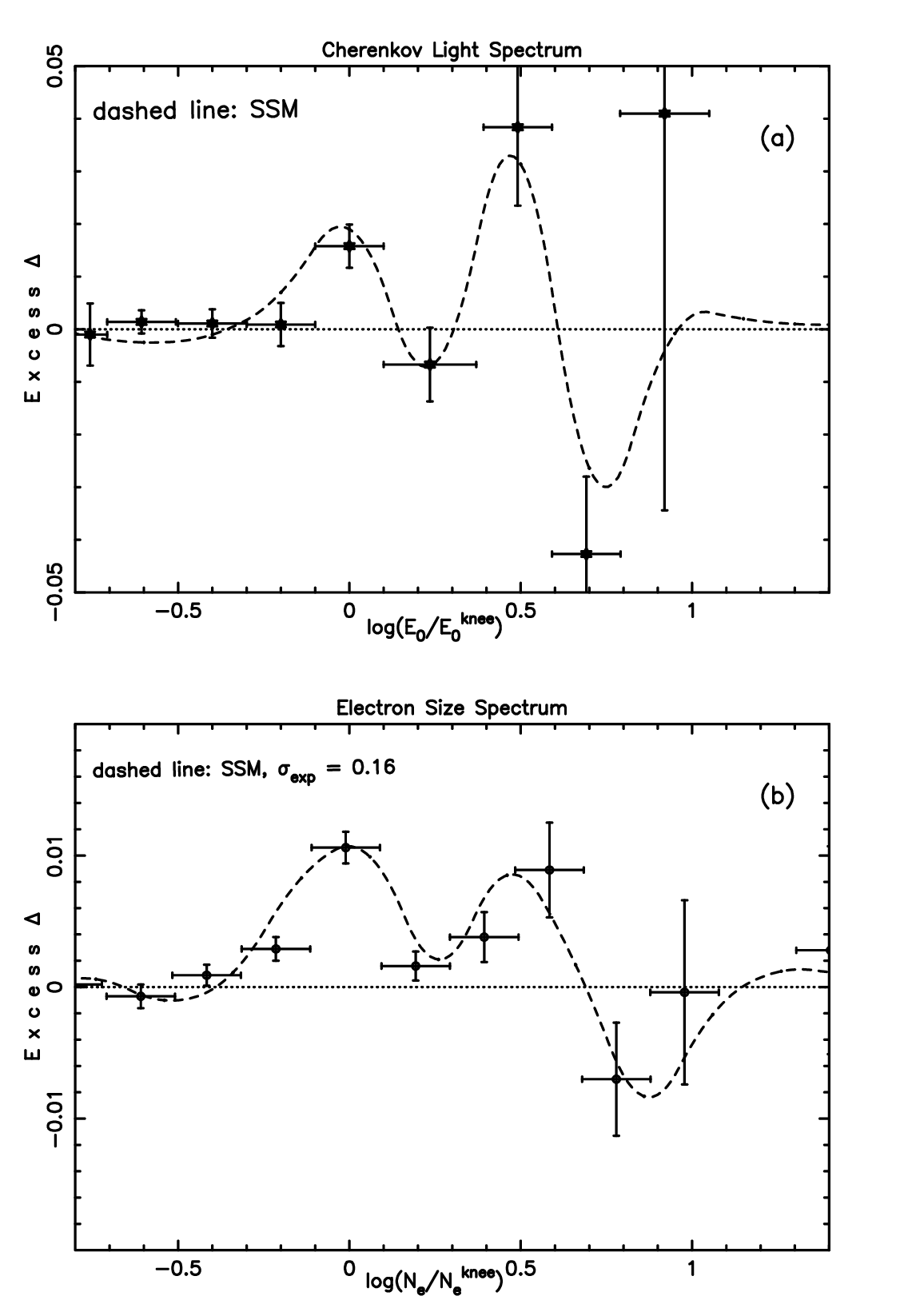}
\end{center}
\caption{Excess over the running mean. The results relate to the
average excess, in $\Delta log(E_0/E_0^{knee})$ and $\Delta
log(N_e/N_e^{knee})$ bins equal to 0.2, for all the world's data. The
upper graph is for the Cherenkov results and the lowest for the
electron size spectrum. The dashed curves for the Cherenkov and electron
spectrum graph are the SSM predictions; that for
Electron Size Spectrum includes an `experimental error' with $\sigma_{exp} = 0.16$.}
\label{fig:fig1}
\end{figure}
The magnitude of sharpness and the existence of the
second intensity peak are important for the problem of the origin
of the knee, because they leave
{\em no room for the Galactic Modulation model} with its smooth and regular
steepening of all the constituent nuclei spectra.  

\subsection{Mass composition}   

The primary mass composition in the knee region is still a matter of
hot debate. The problem is that direct measurements in space do not
yet reach the important PeV region. All the studies of the mass
composition there are indirect and based on the ratios between different
shower components. The range of conclusions is highly disparate;
however there has been progress here in the last few years. 
Most of the experiments now give convergent results and conclude
that the mass composition becomes heavier beyond the knee.

\subsection{Anisotropy} 

The anisotropy of arrival directions can provide 
information about the origin of the knee. The overall situation has
been discussed by us in \cite{elw}. We only
underline that both the amplitude and the 
phase of the first harmonic show sharp changes in the
PeV-region. It is another argument in favour of an astrophysical
origin and against the interaction model of the knee.  

\section{Single Source Model of the Knee (SSM)}

There is a general conjecture, based on energy and theoretical
arguments, that supernova explosions are responsible for the formation
of the cosmic ray energy spectrum below, and possibly even beyond, the
knee. The intensity of their explosions is correlated with the star   
forming regions, whose properties: density and temperature of the
interstellar gas, strength and irregularity of magnetic fields {\em
etc} vary over a wide range. Thus, any kind of averaging over the
range of supernovae would eventually result in a smoothly varying
cosmic-ray spectrum. In 1997 we put forward a model in which the knee
is formed by the explosion of just {\em a single, nearby and recent
supernova} \cite{ew1}. 
\subsection{Shape of the primary energy spectrum}
The shape of the primary energy spectrum in SSM is shown in Figure
2. The spectrum of cosmic rays caused by the explosion of single SN 
protrudes through the smooth background formed by many other 
sources.    
\begin{figure}[htb]
\begin{center}
\includegraphics[height=5cm,width=12cm]{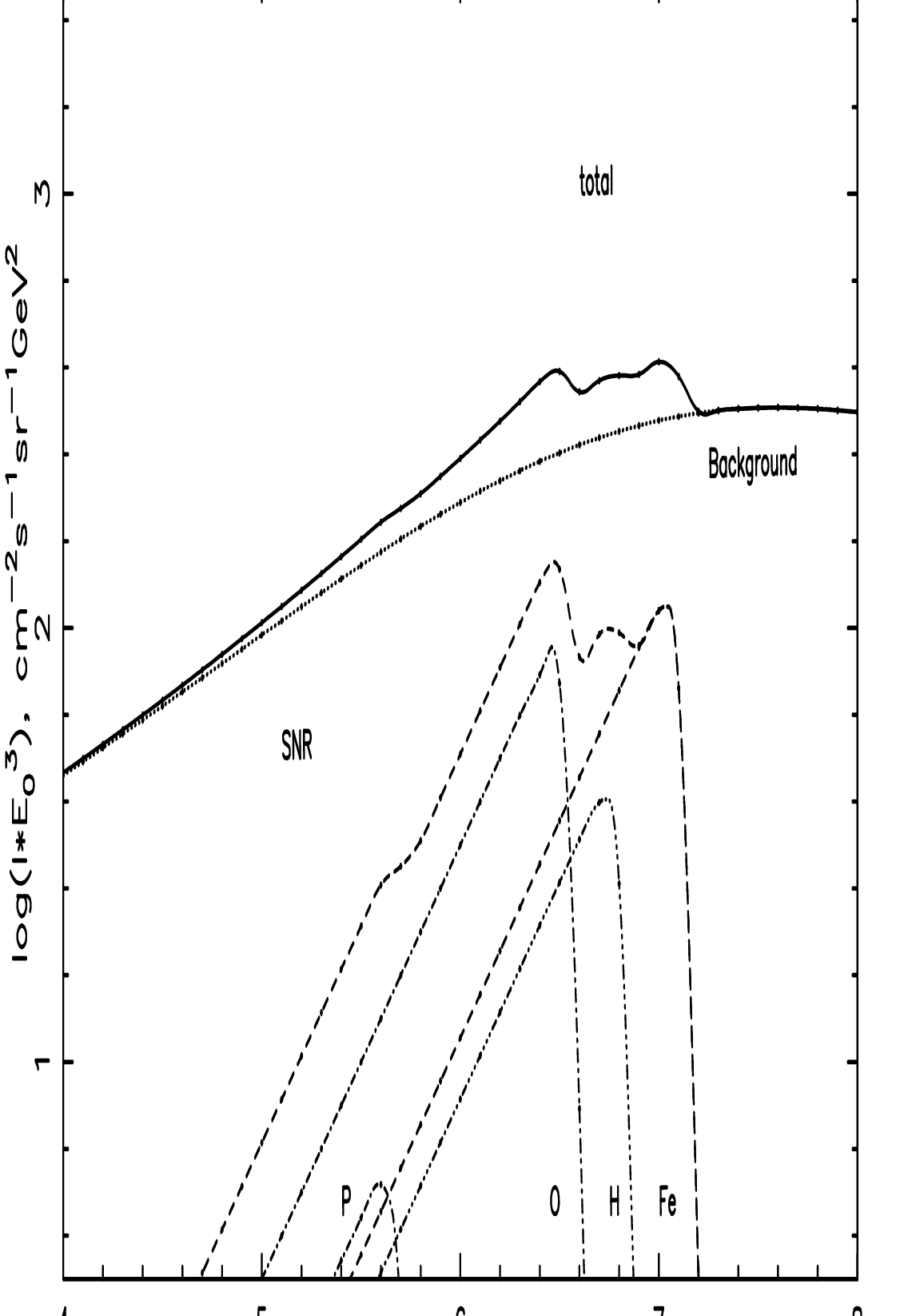}
\end{center}
\caption{The Single Source Model of the primary energy spectrum.
Structure of the spectrum: SNR denotes the contribution from the
single supernova and lines denoted as P, O, H, Fe - the contribution of the constituent spectra from protons, oxygen, heavy ( Ne-S ) and iron nuclei correspondingly. The background spectrum is assumed to be due to many 
(unspecified) sources.}
\label{fig:fig2}
\end{figure}
We attribute the knee to the contribution of oxygen nuclei,
because:\\
(i) its position at 3 PeV corresponds to the theoretical prediction
for oxygen;\\
(ii) its position and intensity stand well at the extrapolation from the
results of direct measurements at lower energies;\\
(iii) it helps to understand the sharpness of the EAS size spectra,
because the nuclei-initiated showers have smaller fluctuations in
their development and the sharp cutoff in the primary energy spectrum
should not be diluted when transferred to the EAS size spectrum.      
\subsection{Mass composition}
(i) If the knee is attributed to oxygen then the second peak is
probably 
associated with iron. This assumption
explains well the existence of the second peak and its separation
from the knee by a factor proportional to the charge (~see SSM
based curves in Figure 1~);\\ 
(ii) in the SSM the primary mass should rise with
energy beyond the knee, which agrees with the results of the
experiments mentioned in subsection 3.2.
\subsection{Anisotropy}
Despite the fact that the source in the SSM is recent and nearby it is certain
that it should not create a very strong anisotropy. If our
interpretation of the mass composition, i.e. the dominance of oxygen
and iron in the peaks is correct, then the peak energies correspond 
to a rigidity of 0.4 PV. The maximum Larmor radius of all
the nuclei at a rigidity of 0.4 PV in the surroundings of the solar
system is about 0.1 pc. The propagation of the cosmic rays from the
source is definitely by diffusion and the anisotropy is therefore determined
just by the gradient of the very local cosmic-ray density. The second factor
which might be important is the location of the solar system with
respect to the shock front. If we are inside it, the cosmic rays are
highly isotropized and even their gradient is not easy to
detect. Perhaps the changes of amplitude and phase of the first
harmonic are the only imprints of the nearby source on
the generally isotropic flux of the cosmic rays at PeV energies. 
\section{On the way to the identification of the single source}
There are not many known single sources which could be classified as 
`nearby and recent' {\em ie} within the range of a few hundred parsecs
and a few hundred thousand years ( Figure 3 ). 

 The first step in the identification of the
source responsible for the knee is to determine whether we are inside
or outside the shock front. We argue that the case when we are 
{\em inside the shock} ( or only just outside ) is
preferable, compared with the opposite case when we are far 
outside it \cite{ew3}.
We remark that this conclusion helps us to understand also the relatively
small amplitude of the anisotropy in the knee region. The typical
propagation of the shock wave from the supernova explosion, taken from 
the Berezhko {\em et al.} calculations \cite{berez}, is shown in
Figure 3 by the dotted line. The sources inside the shock should lie
below this line.
\begin{figure}[htb]
\begin{center}
\includegraphics[height=5cm,width=12cm]{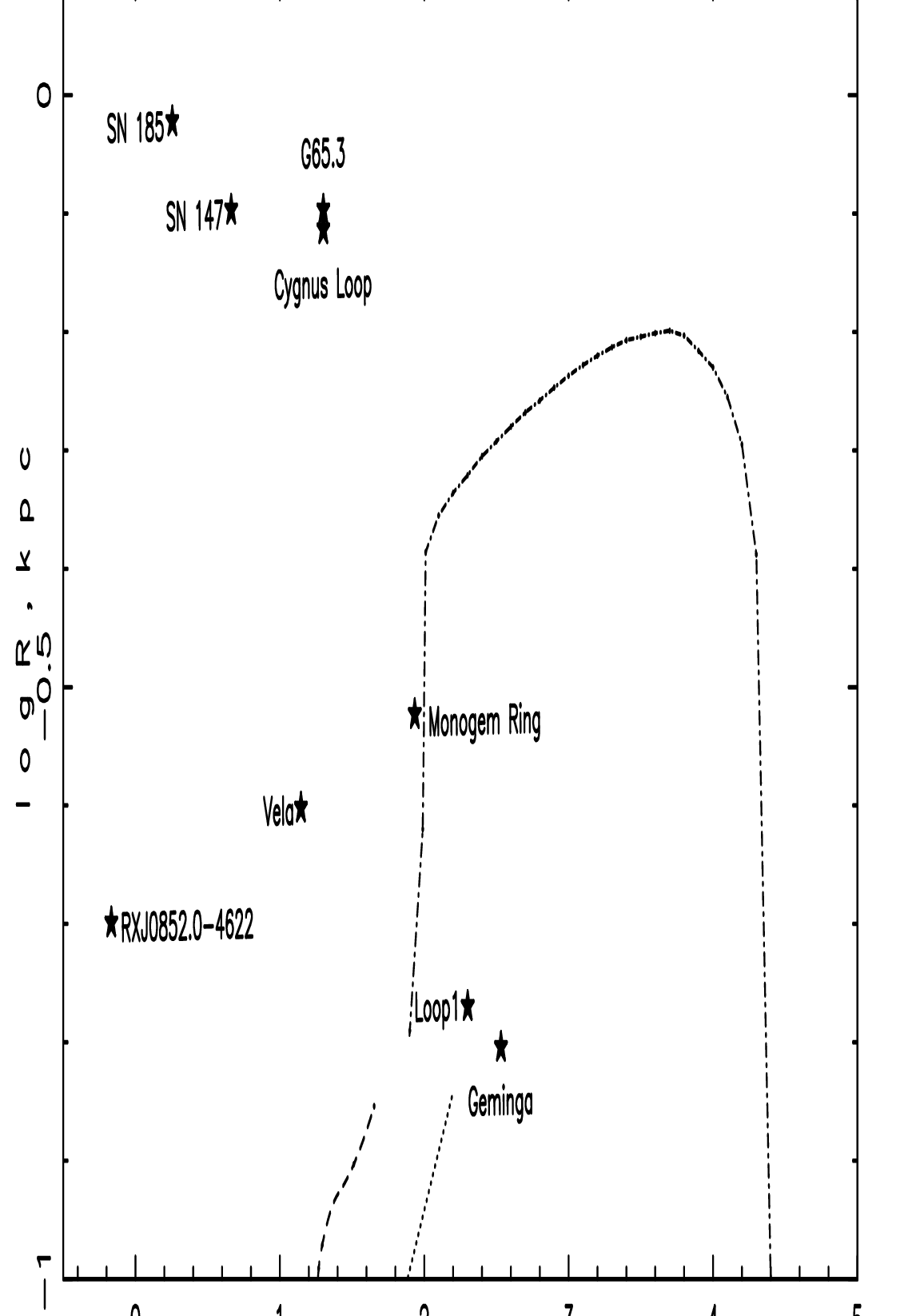}
\end{center}
\caption{Plot of recent and nearby sources, which might be responsible
for the formation of the knee. The dotted line shows the typical
propagation of the shock front. The dashed and dash-dotted lines are
derived from the comparison of the energy contained in cosmic rays for
our Single Source
and the energy content of the cosmic rays accelerated by the supernova
remnant in the calculations of Berezhko {\em et al.}$^5$ and in our
model $^6$. The possible source candidates should lie inside the area 
delimited by these lines.}
\label{fig:fig3}
\end{figure}

Another step in the identification of the source might be based on the
analysis of the energy content of the source. The candidate for the
source could not be too far from the solar system or too close to the
moment of the explosion in order to have enough energy in cosmic rays
and give the requred contribution to the cosmic-ray flux at the knee. 
Comparison of the energy
density in our single source ( Figure 2 ) with the relevant
calculations \cite{berez} and also using our model of the SN explosion
\cite{ew4} indicates that our single source should lie to the right of
the dashed line and inside the area limited by the dash-dotted line in
Figure 3, closer to its left and lower corner.  At the moment, 
the sources which gave birth to Loop I, Monogem Ring and the Geminga
pulsar are the most favorable contenders. However, the problem of the
identification is still with us.
\section{Conclusion}  
The recent results on EAS allow us to conclude that: a)
the knee observed in the cosmic-ray spectrum at about 3 PeV
has an astrophysical origin; b) the sharpness and the fine structure of
the knee rule out the Galactic Modulation Model as the origin of the knee;
c) the most likely model of the knee is the {\em source} model in
which the knee appears as the result of the explosion of a single,
recent, nearby supernova.       

\end{document}